\documentstyle[pra,aps]{revtex}

\input epsf

\newcommand{\be}{\begin{equation}}
\newcommand{\ee}{\end{equation}}
\newcommand{\ba}{\begin{eqnarray}}
\newcommand{\ea}{\end{eqnarray}}
\newcommand{\ord}{{\cal O}}

\begin{document}
\twocolumn[\hsize\textwidth\columnwidth\hsize\csname
@twocolumnfalse\endcsname

\title{Critical states of transient chaos}
\author{Z. Kaufmann$^{1}$, A. N\'emeth$^{1,2}$, and P. Sz\'epfalusy$^{1,3}$}
\address{$^1$Department of Physics of Complex Systems,
E\"otv\"os University,
P.O. Box 32, H-1518 Budapest, Hungary\\
 $^2$Brody Research Center, GE Lighting Tungsram Co.\ Ltd.,
V\'aci \'ut 77, H-1340 Budapest, Hungary\\
 $^3$Research Institute for Solid State Physics and Optics,
P.O. Box 49, H-1525 Budapest, Hungary}
\maketitle

\begin{abstract}%
One-dimensional maps exhibiting transient chaos
and defined on two preimages of the unit interval {[0,1]} are investigated.
It is shown
that such maps have continuously many conditionally invariant
measures $\mu_{\sigma}$ scaling at the fixed point at $x=0$
as $x^{\sigma}$, but smooth elsewhere.
Here $\sigma$ should be smaller than a critical value $\sigma_{c}$
that is related to the spectral properties of the
Frobenius-Perron operator.
The corresponding natural measures are proven to be entirely
concentrated in the fixed point.
\end{abstract}
\vskip2pc] 

\narrowtext

\section{Introduction}

Transient chaos has attracted an increasing interest in the 
last decade due to its
connection to diffusion \cite{GaN,Ga,T96,KlD99}
and chaotic advection \cite{SoKuGi,KT}.
Transiently chaotic behavior develops often in a time period preceding
the convergence of the trajectories to an attractor, or their escape from the
considered region of space as is the case of chaotic scattering.
The length of this time-period depends on the starting point
of the trajectory and is unlimited, there are trajectories (though with
Lebesgue measure zero) that never escape. 
The behavior
of the very long trajectories is governed by the properties of
the maximal invariant set,
the chaotic repeller and the natural measure on it \cite{KG}.
This measure is related to the conditionally invariant measure \cite{PY},
namely the former one is the restriction of the latter one to the repeller
accompanied with a normalization to unity there
(see for reviews \cite{T90,CsGySzT}).
 
Transient chaos is much richer in possibilities than the permanent one.
Regarding e.g. the frequently studied chaotic systems, the 1D maps,
there are rigorous theorems stating that in case of everywhere
expanding maps exhibiting permanent chaos there exists a unique
absolutely continuous invariant measure. However, this is not
any more valid in case of transient chaos for the conditionally
invariant measure \cite{KLNSz}, which in many respects takes over the role
of an invariant measure.
The main purpose of the present paper is to further investigate
this question. It will be shown that one has to distinguish
between normal (non-critical) and critical conditionally
invariant measures. While the first is typically unique, there
are continuously many critical conditionally invariant measures.
The latter ones deserve their name since their corresponding
natural measure is degenerate. Namely, it is non-zero only
on a non-fractal subset of the repeller, on a fixed point 
in case of 1D maps, we are going to study in the present paper.

The map generated by the function $f(x)$ is assumed
to map two subintervals $I_0$ and $I_1$ of $[0,1]$ to the whole $[0,1]$
(see Fig.~1).
It is monotonically increasing in $I_0=[0,\hat x_0]$
and decreasing in $I_1=[\hat x_1,0]$,
$f(0)=f(1)=0$ and $f(\hat x_0)=f(\hat x_1)=1$.
The value of $f$ is undefined in $(\hat x_0,\hat x_1)$, and we consider
the trajectory being in this interval to escape in the next iteration.
We assume $f$ is smooth
and hyperbolic ($1<|f'(x)|<\infty$) in $I_0$ and $I_1$,
or we allow singular behavior with
infinite slope in $\hat x_0$, $\hat x_1$, and $x=1$.

\begin{figure}
\epsfbox{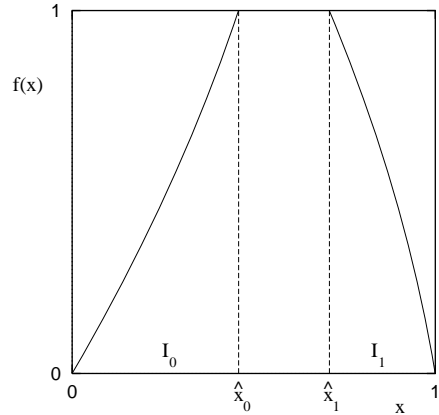}
\caption{Schematic plot of the map $f(x)$.}
\end{figure}

Instead of treating the Frobenius-Perron operator for the density
$P^{(k)}(x)$ we deal with the measure 
$\mu^{(k)}(x)=\mu^{(k)}([0,x])$, where 
$\mu^{(k)}([x_1,x_2])=\int_{x_1}^{x_2} P^{(k)}(x)\,dx$.
Note that $\mu^{(k)}(x)$
is a monotonically increasing function. 
The upper index refers to the discrete time.
The equation of time evolution for the measure 
can be written as
\be
\mu^{(k+1)}(x)=T\mu^{(k)}(x)\equiv T_0\mu^{(k)}(x)+T_1\mu^{(k)}(x)\;,\label{t}
\ee
where the contributions of the two branches are
\ba
T_0\mu^{(k)}(x)&=&\mu^{(k)}(f_0^{-1}(x))\;,\;\;\nonumber\\
T_1\mu^{(k)}(x)&=&\mu^{(k)}(1)-\mu^{(k)}(f_1^{-1}(x))\;.\label{ti}
\ea
$f_0^{-1}(x)$ ($f_1^{-1}(x)$) denotes the lower (upper) branch
of the inverse of $f(x)$.
Since a portion of the trajectories escapes in every step,
normalization is necessary to ensure that the iteration
converges to a certain measure \cite{PY,T90,CsGySzT},
which is then an eigenfunction of $T$, namely
\be
T\mu(x)=e^{-\kappa}\mu(x)\;.\label{cim}
\ee
The measure $\mu$ is called conditionally invariant measure
(the notation $\mu$ without an upper index always refers to that
in the present paper), and $\kappa$ is the escape rate.
The afore mentioned definition of the natural measure $\nu$ yields its
connection to $\mu$.
Namely, the natural measure \cite{KG} of a non-fractal set $A$ ($\mu(A)>0$)
is given by
\be
\nu(A)=\lim_{n\rightarrow\infty}
\frac{\mu(A\cap f^{-n}[0,1])}{\mu(f^{-n}[0,1])}\;.
\ee

The paper is organized as follows.
The infinity of the coexisting conditionally invariant measures
and the general condition for criticality are presented in Section 2.
To get a deeper understanding of the conditionally invariant measures
we study the spectrum of
the Frobenius-Perron operator in Section 3.
In Section 4 we show how maps with critical state are connected to each other
by singular conjugation,
thereby the singular measures are brought into connection with 
nonsingular ones.
In Section 5 it is shown that the natural measures corresponding to critical
states are fully concentrated to the fixed point at $x=0$, which is
the main property of criticality.
Section 6 is devoted to demonstrate some of the results on examples
with further discussion.

\section{Conditionally invariant measures}

First we assume that $f$ is nonsingular in $I_0$ and $I_1$, and
in the second part of this section we shall study the case when the map
may be singular with infinite slope in $x=\hat x_0, x=\hat x_1$ and/or
$x=1$.
We shall see that even in the first case there are continuously many
conditionally invariant measures $\mu_\sigma$ that have different
power law behavior $\mu_\sigma\sim x^\sigma$ with $\sigma>0$ at $x=0$.
In order to show this
we study what a measure do we get asymptotically
when we start with an initial measure $\mu^{(0)}$ that is smooth but
scales as $\mu^{(0)}(x)\approx a x^\sigma$ for $x\ll 1$.
(The simplest possibility is to choose $\mu^{(0)}=x^\sigma$, which will be used
in the numerical calculations.
Note that $\sigma=1$ corresponds to the Lebesgue measure.)
The escape rate obtained in the asymptotics shall be denoted by
$\kappa_\sigma$.
First we study the action of the terms $T_i$ of $T$ on $\mu^{(0)}$ separately
(see Eqs.\ (\ref{t}),(\ref{ti})).
It can easily be seen that the leading term of $T_0 \mu^{(0)}$ is
\be
T_0 a x^\sigma\approx a e^{-\lambda_0\sigma}x^\sigma
\;\;\;\mbox{if}\;\;x\ll 1\;,\label{t0si0}
\ee
where $\lambda_0=\log(f'(0))$ is the local Liapunov exponent at the fixed
point.
On the other hand, since $T_1$ does not take values of $\mu^{(0)}$
from the vicinity of $x=0$,
$T_1 \mu^{(0)}$ is smooth with linear behavior
\be
T_1\mu^{(0)}\approx b x\;\;\;\mbox{if}\;\;x\ll 1\;.\label{t1si0}
\ee
Therefore $T_0\mu^{(0)}$ dominates in $T_0\mu^{(0)}+T_1\mu^{(0)}$ if
$\sigma<1$,
and $T_0\mu^{(0)}$ and $T_1\mu^{(0)}$ both scale linearly
near $x=0$ if $\sigma=1$.
So the scaling of $\mu^{(0)}$ at $x=0$ is retained if $\sigma\le 1$.
That is why we expect that the conditionally invariant measure
and the corresponding escape rate may depend on $\sigma$.

Turning to the asymptotics for large time we rewrite the $n$-th iterate as
\be
T^n\mu^{(0)}=T_0^n\mu^{(0)}+\sum_{k=1}^nT^{k-1}T_1T_0^{n-k}\mu^{(0)}\;.
\label{tn}
\ee
It can be seen from Eq.\ (\ref{t0si0}) that the first term on the r.\ h.\ s.\
of (\ref{tn}) gives
a contribution $a e^{-\lambda_0\sigma n}x^\sigma$ for small $x$.
The similar factor in the sum can also be estimated as
$T_0^{n-k}\mu^{(0)}\approx a e^{-\lambda_0\sigma(n-k)}x^\sigma$.
Acting on this function by $T_1$ a function is created that is proportional
to $x$ in the vicinity of $x=0$, similarly to (\ref{t1si0}).
Therefore $T_1T_0^{n-k}\mu^{(0)}$ yields an
asymptotics $e^{-\kappa_1 k}$ for large $k$ under the action of $T^{k-1}$.
Since for large $n$ at least one of $n-k$ and $k$ is large we obtain
\be
T^n\mu^{(0)}\approx a e^{-\lambda_0\sigma n}x^\sigma
+\sum_{k=1}^n\ord\left(e^{-\lambda_0\sigma(n-k)}e^{-\kappa_1 k}x\right)\;.
\ee
Consequently, in case $\lambda_0\sigma<\kappa_1$ ($\lambda_0\sigma>\kappa_1$)
the first (last) term dominates for large $n$ and small $x$ and
$T^n \mu^{(0)}$ behaves asymptotically as $e^{-\lambda_0\sigma n}$
($e^{-\kappa_1n}$).
That means, there is a critical value $\sigma_c=\kappa_1/\lambda_0$ such that
for every $\sigma<\sigma_c$
in the limit $n\rightarrow\infty$ with normalization in each step
we obtain a conditionally invariant measure $\mu_\sigma$ with leading term
proportional to $x^\sigma$ at $x=0$,
while in case $\sigma>\sigma_c$ we obtain $\mu_1$.
It is easy to see applying $T$ on $\mu_1$
that $\kappa_1<\lambda_0$, i.\ e.\ $\sigma_c<1$.
The corresponding escape rates are
\ba
\kappa_\sigma&=&\lambda_0\sigma\;\;\;
\mbox{if}\;\;\sigma<\sigma_c\;,\label{ks1}\\
\kappa_\sigma&=&\kappa_1\;\;\;\mbox{if}\;\;\sigma>\sigma_c\;.\label{ks2}
\ea
So the escape rate in case $\sigma<\sigma_c$ is determined by the slope taken
at the fixed point $x=0$.
We consider the system to be critical with respect to $\mu_\sigma$
if $\sigma<\sigma_c$ since
the density of the corresponding natural measure is
a Dirac delta function located at the origin, as will be shown in Section 4.
Deeper understanding of Eqs.\ (\ref{ks1}), (\ref{ks2})
shall be reached in the next section by studying the spectrum of $T$.

In the second part of this section we allow
singularities of $f$ at the maximum points $\hat x_i=f_i^{-1}(1)$ with $i=0,1$
and/or at $x=1$.
Namely, the inverse branches behave as
\ba
f_i^{-1}(x)&\approx&\hat x_i + B_i(1-x)^\psi\;\;\;
\mbox{if}\;\; 1-x\ll 1\;,\label{f1}\\
f_1^{-1}(x)&\approx&1 - C x^\omega\;\;\;
\mbox{if}\;\; x\ll 1\;,\label{f0}
\ea
where $\psi\ge 1$ and $\omega\ge 1$.
If $\psi>1$ ($\omega>1$) the map has infinite slope at the maximum
points $\hat x_0, \hat x_1$ (at the point $x=1$).
Studying the effect of one iteration on a monotonic function $\chi(x)$
that is smooth in $(0,1)$ and obeys scaling
$\chi(x)\approx a x^\sigma$ at $x=0$
we see that
Eq.\ (\ref{t0si0}) remains valid.
$T_0\chi(x)$ and $T_1\chi(x)$ have leading terms
with exponent $\psi$ at $x=1$,
\be
T_i\chi(x)\approx T_i\chi(1)-b_i(1-x)^\psi\;,\;\; i=0,1\;\;\;
\mbox{if}\;\; 1-x\ll 1\;.\label{p}
\ee
If $\chi(x)$ scales as $\chi(x)\approx \chi(1)-b(1-x)^\psi$ near $x=1$
then acting by $T_1$ on it the result scales as
\be
T_1\chi(x)\approx c x^\beta\;\; \mbox{with}\;\beta=\psi\omega\;\;\;
\mbox{if}\;\; x\ll 1\;.\label{b}
\ee
Consequently, starting with a $\chi(x)$ that satisfies
\ba
\chi(x)&\approx& a x^\sigma\;\;\; \mbox{if}\;\; x\ll 1\;,\label{p1}\\
\chi(x)&\approx& \chi(1)-b(1-x)^\psi\;\;\; \mbox{if}\;\; 1-x\ll
1\label{p2}
\ea
$T\chi$ retains these properties if $\sigma\le \beta$.

We start with a $\mu^{(0)}$ in a class given by Eqs.\ (\ref{p1}), (\ref{p2})
and investigate Eq.\ (\ref{tn}) similarly to the case of maps nonsingular in
$I_0,I_1$, which maps correspond to the case $\beta=1$.
By Eq.\ (\ref{t0si0}) we obtain again
that $T_0^n\mu^{(0)}\approx a e^{-\lambda_0\sigma n}x^\sigma$
and $T_0^{n-k}\mu^{(0)}\approx a e^{-\lambda_0\sigma(n-k)}x^\sigma$ near $x=0$.
According to Eqs.\ (\ref{p}) and (\ref{b})
$T_1 T_0^{n-k}\mu^{(0)}$ belongs to the class of functions defined by
Eqs.\ (\ref{p1}) (\ref{p2}) with $\sigma=\beta$.
Its iterates by $T^{k-1}$ decay proportionally to $e^{-\kappa_\beta k}$ for
large $k$.
Since for large $n$ either $k$ or $n-k$ is large, we finally estimate
$T^n\mu^{(0)}$ using Eq.\ (\ref{tn}) as
\be
T^n\mu^{(0)}\approx a e^{-\lambda_0\sigma n}x^\sigma
+\sum_{k=1}^n\ord\left(e^{-\lambda_0\sigma(n-k)}e^{-\kappa_\beta k}x\right)\;.
\ee
That means, the border value of $\sigma$ is now
$\sigma_c=\kappa_\beta/\lambda_0$.
In case $\sigma<\sigma_c$ we obtain a conditionally invariant measure
$\mu_\sigma$ belonging to the class of functions given by Eqs.\ (\ref{p1}),
(\ref{p2}),
while in case $\sigma>\sigma_c$ we obtain the conditionally invariant
measure $\mu_\beta$, which belongs to the class given by (\ref{p1}), (\ref{p2})
with $\sigma=\beta$.
Applying $T$ on $\mu_\beta$ one can easily see
that $\kappa_\beta<\lambda_0$, i.\ e.\ $\sigma_c<\beta$.
The corresponding escape rates are
\ba
\kappa_\sigma&=&\lambda_0\sigma\;\;\;
\mbox{if}\;\;\sigma<\sigma_c\;,\label{ks3}\\
\kappa_\sigma&=&\kappa_\beta\;\;\;\mbox{if}\;\;\sigma>\sigma_c\;.\label{ks4}
\ea
Again the escape rate in case $\sigma<\sigma_c$ is determined alone by the
slope of the map at $x=0$ and the measure belonging to such a $\sigma$
represents a critical state.
Let us emphasize that while there is a continuum infinity of
critical conditionally invariant measures
the noncritical one is unique.

It seems to be impossible to determine the full basin of attraction of the
conditionally invariant measures.
In the class of functions that are monotonic and smooth in $(0,1)$
those belong to the basin of attraction of $\mu_\sigma$ with $\sigma<\sigma_c$
that
\begin{itemize}
\item scale as $a x^\sigma$ at $x=0$
and not slower than $\mu^{(0)}(1)-b(1-x)^{\sigma/\omega}$ at $x=1$,
\item or scale faster than $a x^\sigma$ at $x=0$
and scale as $\mu^{(0)}(1)-b(1-x)^{\sigma/\omega}$ at $x=1$.
\end{itemize}
The basin of attraction of $\mu_\beta$ consists of the functions
that scale faster than $a x^\sigma_c$ at $x=0$
and faster than $\mu^{(0)}(1)-b(1-x)^{\sigma_c/\omega}$ at $x=1$.

Note that the possible singular behavior of the noncritical conditionally
invariant measure is determined completely by the map.
The behavior of critical conditionally invariant measures on the right hand
side is also determined by the map.
By this reason we classify these critical measures by the behavior near $x=0$.
Their leading term at $x=0$ is analytic when $\sigma$ is integer.
The number of such measures is $[\sigma_c]$, where [ ] denotes integer part.
If $\psi=\omega=1$ then $[\sigma_c]=0$.

\section{Eigenvalue spectrum}

The conditionally invariant measures obtained in the previous 
section are particular eigenfunctions
of the operator $T$ (see Eqs.\ (\ref{t},\ref{ti})),
namely, they are monotonous (positive definite)
functions. To get further insight into the appearance
of the upper value $\sigma_c$ of the parameter $\sigma$ specifying the
critical measures we study more general eigenfunctions of the operator
$T$.
We allow that $f$ may have
singularity at $x=\hat x_0$, $x=\hat x_1$ and/or in $x=1$,
as in the second part of the previous section.
However, for sake of simplicity here we also assume that
the inverse branches of the map are analytic,
so $\psi$ and $\omega$ in Eqs.\ (\ref{f1}), (\ref{f0}) (and thereby $\beta$ in
(\ref{b})) are integers.
We also assume, as it is typical, that there is a discrete spectrum of the
Frobenius-Perron operator in the space of analytic functions.
This has been proved for certain one-parameter families of maps
\cite{GySz/J,GySz/A,Ke,CsGySzT}.

We shall see that for any value of $\sigma$
an expansion in terms of the basis functions $T_0 x^{\sigma+n}$ and
$x^{\beta+n}$ with $n=0,1,\ldots$ is
convenient for the search of eigenfunctions. Therefore we start from the
form
\be
\phi=\sum_{n=0}^{N(\sigma)-1}c_n T_0 x^{\sigma+n}
+\sum_{n=0}^\infty d_n x^{\beta+n}\;,\label{e}
\ee
where $N(\sigma)=\beta-\sigma$ if $\sigma$ is integer,
and $N(\sigma)=\infty$ otherwise.
The limitation by $N(\sigma)$ is necessary if $\sigma$ is integer, since
$T_0 x^{\sigma+n}$
can be expanded on the basis functions $x^{\beta+l}$ if
$\sigma+n$ is an integer greater or equal to $\beta$. Note that
\be
T_0 x^{\sigma+n}=\sum_{m=0}^\infty g_{mn}x^{\sigma+m}\;.
\ee
It also follows that
\ba
g_{mn}&=&e^{-\lambda_0(\sigma+m)}\;\;\; \mbox{if}\;\; m=n\;,\label{gd}\\
g_{mn}&=&0\;\;\; \mbox{if}\;\; m<n\;,\label{g}
\ea
and the basis functions in the first sum of Eq.\ (\ref{e})
are transformed by $T_0$ as
\be
T_0 T_0 x^{\sigma+n}=\sum_{m=0}^\infty g_{mn}T_0 x^{\sigma+m}\;.\label{e0}
\ee
The transformation by $T_1$ can be obtained similarly to Eq.\ (\ref{b}),
\be
T_1 T_0 x^{\sigma+n}=\sum_{m=0}^\infty H_{mn}
x^{\beta+m}\;.\label{e1}
\ee
Clearly, the basis functions in the second sum of Eq.\ (\ref{e})
are transformed by $T$ in the way
\be
T x^{\beta+n}=\sum_{m=0}^\infty Q_{mn} x^{\beta+m}\;.\label{e2}
\ee
As seen from Eqs.\ (\ref{e0}), (\ref{e1}) and (\ref{e2}) the iteration of the
vectors
$\bf c,d$ formed from the expansion coefficients in Eq.\ (\ref{e}) under the
action of $T$ can be described as
\be
T\left(\begin{array}{c}\bf c\\\bf d\end{array}\right)=
\left(\begin{array}{cc}\bf G&\bf 0\\ \bf H&\bf Q\end{array}\right)
\left(\begin{array}{c}\bf c\\\bf d\end{array}\right)\;,
\ee
where $\bf G$ is a matrix constructed from the coefficients $g_{mn}$ but with
truncation to size $N(\sigma)\times N(\sigma)$ if $\sigma$ is integer.

In case ${\bf c}=0$ only $\bf Q$ is in effect, so the eigenvalue problem
yields the eigenvalues $\Lambda_{\beta,n}$ of $\bf Q$ and corresponding
eigenfunctions $\phi_{\beta,n}$, whose expansion starts with $x^\beta$.
On the other hand, since $\bf G$ is a triangular matrix
its leading eigenvalue is $e^{-\lambda_0\sigma}$
(see Eqs.\ (\ref{gd}), (\ref{g}))
that belongs to an eigenvector denoted by ${\bf c}_\sigma$.
Finally, an eigenfunction $\phi_\sigma$ of $T$ with an $x^\sigma$ scaling at
$x=0$ can be obtained
in the form of (\ref{e}) with eigenvalue $\Lambda_\sigma$
and coefficients
\be
{\bf c=c_\sigma}\;,\;\;
{\bf d}=({\bf g}_{00}-{\bf Q})^{-1}{\bf H c_\sigma}\;,
\ee
where $\Lambda_\sigma={\bf g}_{00}=e^{-\lambda_0\sigma}$,
except the special cases when $\sigma$
is an integer greater or equal to $\beta$
or $\Lambda_\sigma$ coincides with an eigenvalue
$\Lambda_{\beta,n}$ of $\bf Q$.
Thereby we have obtained besides the assumed discrete spectrum of analytic
eigenfunctions an almost continuous spectrum of $T$.
Most of these eigenfunctions have nonanalyticity due to noninteger value of
$\sigma$.

The next important question is, how the conditionally invariant measures can be
selected from these eigenfunctions.
A measure should be nonnegative for any set, so the condition is that
$\phi(x)$ should be monotonous.
This is ensured when it can be generated starting from a monotonous
$\phi^{(0)}(x)$
as the limit of iteration $T^n \phi^{(0)}(x)$, normalizing it in each step.
We can study this condition using the above results.
After one iteration of a general monotonic $\phi^{(0)}$ with $\ord(x^\sigma)$
scaling at $x=0$ the iterate $T\phi^{(0)}$ is a linear combination of
the basis functions $T_0 x^{\sigma+n}$ and $x^{\beta+n}$ with $n\ge 0$.
Therefore the limit of infinite iterations yields the eigenfunction
that has the largest eigenvalue among the eigenfunctions that
can be expanded on this basis.
That one is the eigenfunction $\phi_\sigma$ if
$\Lambda_\sigma=e^{-\lambda_0\sigma}>\Lambda_{\beta,0}$,
i.\ e.\ $\sigma<-\log(\Lambda_{\beta,0})/\lambda_0$.
On the other hand, when starting with an initial $\phi^{(0)}$ with
$\ord(x^\beta)$ scaling at $x=0$
we do not get terms with any smaller exponent, thereby we obtain
the eigenfunction $\phi_{\beta,0}$.
Therefore the correspondence between these eigenfunctions and the conditionally
invariant measures and between the eigenvalues and escape rates
can be described as
\ba
\phi_\sigma=\mu_\sigma&\;,\;\;& \Lambda_\sigma=e^{-\kappa_\sigma}\;\;\;
\mbox{if}\;\; \sigma<\sigma_c=\kappa_\beta/\lambda_0\;,\label{kl}\\
\phi_{\beta,0}=\mu_\beta&\;,\;\;& \Lambda_{\beta,0}=e^{-\kappa_\beta}\;.
\ea

The spectrum of the Frobenius-Perron operator allowing singular
eigenfunctions has been studied for piece-wise linear maps by
MacKernan and Nicolis \cite{MN}.
Except the tent map, they considered eigenfunctions singular at internal
points of the interval.
They pointed out the existence of the
continuous parts of the spectrum, but did not raise the question 
of the possible monotonous property of the eigenfunctions for a
region of eigenvalues, which has been our main concern here.

Finally we note that singular eigenfunctions of the generalized
Frobenius-Perron operator have been of importance in the
thermodynamic formalism to describe phase transition like 
phenomena, where the "temperature" has played the role of the 
control parameter \cite{FPT,SzTV}.

\section{Conjugation}

It can 
be seen that all critical systems can be brought to 
the same form by the application of smooth
conjugation. For this purpose a conjugation function $u$ has 
to be introduced, which is smooth everywhere except in $x=0$ and $x=1$,
where it may be singular. These singularities can be characterized
by the exponents $\eta$ and $\alpha$:
\begin{eqnarray}
u(x) & \approx & x^\eta \;\;\; \mbox{if}\;\; x \ll 1\;, \label{u0} \\
u(x) & \approx & 1-(1-x)^\alpha\;\;\; \mbox{if}\;\; 1-x \ll 1\;. \label{u1} 
\end{eqnarray} 
By definition, the conjugation transforms the map and the 
measure in the following way: 
\begin{eqnarray}
\tilde{f}^{-1}_i(x) & = & u \left( f^{-1}_i \left( u^{-1}(x) \right) 
\right)\;, \label{ConjugationMap} \\ 
\tilde{\mu}(x)      & = & \mu \left( u^{-1}(x) \right)\;. 
\label{ConjugationMeas} 
\end{eqnarray}
To see how the conjugation
changes the exponents important from the point of view of criticality, 
transformation (\ref{ConjugationMeas}) has to be applied to the 
conditionally invariant measure, and transformation (\ref{ConjugationMap})
to the branches of the inverse map. 
The conditionally invariant measure $\mu$ is
in the class of functions given by Eqs.\ (\ref{p1},\ref{p2}) and
the branches of the inverse map are described in Eqs.\ (\ref{f1},\ref{f0}).
The conjugation results in the following transformation rules for the 
characterizing exponents:
\begin{eqnarray}
\label{conjug}
\tilde{\psi}      & = & \frac{\psi}{\alpha}       \;,\\ 
\tilde{\omega}    & = & \alpha \frac{\omega}{\eta}\;,\\ 
\tilde{\lambda}_0 & = & \lambda_0  \eta           \;,\\
\tilde{\sigma}    & = & \frac{\sigma}{\eta}\;.
\end{eqnarray}
{}From these transformation rules it is clearly seen that by the
application of the appropriately chosen $u$ any two of the three 
quantities 
$\psi$, $\omega$ and $\sigma$ can be set to unity. This means all critical 
systems 
can be brought to the same form, which shows that criticality is the same, 
regardless 
it is caused by the singular measure or the singularity of the map 
in $x=\hat x_0, x=\hat x_1$ and $x=1$.

It is worth noting that any conditionally invariant measure 
can be chosen 
as conjugating function. The conjugation in this case results in the 
equivalent map, which has the Lebesgue measure as a conditionally
invariant one. 
Such maps will be called Lebesgue maps in the following.
The equivalent Lebesgue map will be denoted by 
$\tilde{f}^{(\sigma)}$, if the conditionally invariant measure chosen 
for the conjugation is $\mu_\sigma$, i.e. the measure decaying at 
$x=0$ with the exponent $\sigma$.
Naturally, the Lebesgue measure is not singular in 
$x=0$, so $\tilde{\sigma}=1$. Moreover, since the conditionally invariant 
measure is asymptotically proportional to $(1-x)^\psi$ in $x=1$, the 
conjugation sets the exponent $\tilde{\psi}$ to unity, too, so after the 
conjugation both $\tilde{\sigma}$ and $\tilde{\psi}$ are equal to unity. 
However, if criticality is in
existence, i.e. $\sigma<\sigma_c$, than $\tilde{\omega}$ will be 
greater than one, i.e. the equivalent Lebesgue map is singular in $x=1$. 
It is obvious that the map has as many equivalent Lebesgue maps as
conditionally invariant measures.

\section{Properties of the natural measure}

During the investigations of the piecewise parabolic map it was found that 
the natural measure of the 
fixed point at $x=0$ is positive when the Lebesgue-measure as initial 
measure was iterated \cite{NSz}.
Numerical results suggested, and
later it was supported by analytical considerations that this 
measure is not only positive 
but is equal to unity \cite{NSz,LSz}.
We shall prove here that this phenomenon is quite a general property:
for any 
map with critical conditionally invariant measure the natural measure of 
the fixed point at $x=0$ is equal to unity. 

Since $f^{-1}_0$ has a finite slope $e^{\lambda_0}$ at $x=0$,
\begin{equation}
\label{CSeries}
C_n(x) = \frac{f^{-n}_0(x)}{e^{-\lambda_0 n}x} 
\stackrel{n \rightarrow \infty}{\longrightarrow} C_{\infty}(x)\;,
\end{equation}
where $0<C_{\infty}(x) < \infty$. Furthermore, the critical conditionally 
invariant measure $\mu$ is asymptotically proportional to $x^{\sigma}$,
that is 
\be
m(x) = \frac{\mu(x)}{x^{\sigma}} 
\stackrel{x \rightarrow 0}{\longrightarrow} M\;, \label{CondMeasure}
\ee
where $\sigma<\sigma_c$ and $M$ is finite and positive. 
We introduce the following notation for the set of the 
preimages of the unit interval $I=I^{(0)}_0=[0,1]$. The first two preimage 
intervals are $I^{(1)}_0=f^{-1}_0(I)$ and $I^{(1)}_1=f^{-1}_1(I)$. 
Similarly, the $(n+1)$-th preimages can be generated from the \mbox{$n$-th}
ones as $I^{(n+1)}_i=f^{-1}_0(I^{(n)}_i)$ and 
$I^{(n+1)}_{2^n+i}=f^{-1}_1(I^{(n)}_i)$. 
The set of all the $n$-th preimages of $I$ is denoted by 
$I^{(n)} = \bigcup_{i=0}^{2^n-1} I^{(n)}_i$.
We want to determine the natural invariant measure of a single point,
the fixed point located at $x=0$, a 
series of intervals containing this point must be found, whose limit is 
the fixed point itself.
The natural measure of the fixed point is equal to
the limit of the series of the natural measures of these intervals. 
The series of the leftmost intervals of the $k$-th interval sets is an 
appropriate and convenient choice, so the natural measure of $x=0$ is 
\begin{equation} 
\label{NaturalMeasure2}
\nu(\{0\}) = \lim_{k \rightarrow \infty} \nu(I^{(k)}_0) =
\lim_{k \rightarrow \infty}\lim_{n \rightarrow \infty}
\frac{\mu\left(I^{(k)}_0 \cap I^{(n)}\right)}{\mu\left(I^{(n)}\right)}\;.
\end{equation}
Since $\mu\left(I^{(k)}_0 \cap I^{(n)}\right)>\mu\left(I^{(n)}_0 
\right)$, $k<n$,
we estimate the natural measure from below by keeping only 
the leftmost interval. From the criticality and
Equations (\ref{CondMeasure}) and (\ref{CSeries}) follows 
\begin{eqnarray}
\label{MeasureInx=0}
\nu(\{0\}) &\geq& \lim_{n \rightarrow \infty} 
\frac{\mu(f^{-n}_0(1))}{\mu\left(I^{(n)}\right)} \nonumber \\ &=& 
\lim_{n \rightarrow \infty} \frac{ m\left(C_n(1)e^{-\lambda_0 n} \right) 
\cdot C_n(1)^{\sigma}e^{-\lambda_0 n \sigma}} {e^{-\kappa n}}\nonumber\\
&=& M \cdot C_{\infty}(1)^\sigma > 0\;,
\end{eqnarray}
so the positive natural measure of the fixed point is proven.

It can also be shown that this measure is equal to unity. For this 
purpose the features of the conjugation to an equivalent Lebesgue 
map have to be used. Let us choose as the conjugating function the 
non-critical conditionally invariant measure $\mu_\beta$, 
where $\beta = \psi \omega$. The conjugation results in the map 
$\tilde{f}^{(\beta)}$, which is characterized by exponents
$\tilde{\psi}=1$ and $\tilde{\omega}=\beta$ . The index
$(\beta)$ in the following will be omitted. 
Any $\mu_\sigma$ conditionally invariant measure transforms into
$\tilde{\mu}_{\tilde{\sigma}}$, where $\tilde{\sigma}=\sigma/\beta$.
Neither $\mu_\beta$ nor its conjugated pair, the $\tilde{\mu}_1$ 
Lebesgue measure are critical, so $\tilde{\sigma}<\tilde{\sigma}_c<1$
must hold for any critical $\tilde{\mu}_{\tilde{\sigma}}$ measure. 
Since $\tilde{\mu}_1$ is Lebesgue measure, the
$\tilde{\ell}\left(\tilde{I}^{(n)}\right)$ total length of the $n$-th 
preimage set of the unit interval $I$ is equal to its measure with 
respect to $\tilde{\mu}_1$. This fact makes the exact determination 
of $\tilde{\ell}\left(\tilde{I}^{(n)}\right)$ possible. 
Since $\mu( f^{-n}(A)) = T^{n}\mu(A) = e^{-\kappa n}\mu(A)$
for any set $A \subseteq I$ and $\mu$ conditionally invariant measure
\begin{equation}
\label{CylinderSetLength}
\tilde{\ell}\left(\tilde{I}^{(n)}\right) = 
\tilde{\mu}_1 \left( \tilde{f}^{-n}(\tilde{I}) \right) = 
e^{-\tilde{\kappa}_1 n} = e^{-\kappa_\beta n}
\end{equation}
holds. Now we can calculate the natural measure concentrated in the 
fixed point at $x=0$ for $\tilde{\mu}_{\tilde{\sigma}}$, 
$\sigma<\sigma_c$ critical conditionally invariant measures. 
For this purpose we can use Eq.\ (\ref{NaturalMeasure2}). 
Since, provided that $k<n$, 
$\tilde{I}^{(k)}_0 = [0,\tilde{f}^{-k}_0(1)]$ and 
$\tilde{I}^{(n)} = [0,\tilde{f}^{-k}_0(1)]\cap \tilde{I}^{(n)} \cup 
[\tilde{f}^{-k}_0(1),1] \cap \tilde{I}^{(n)}$, 
we can write that 
\ba
\label{ConjugNuin0}
\tilde{\nu}_{\tilde{\sigma}}(\{0\}) &=&\lim_{k \rightarrow \infty}
\lim_{n \rightarrow \infty} 
\frac{ \tilde{\mu}_{\tilde{\sigma}}( \tilde{I}_0^{(k)} \cap
\tilde{I}^{(n)})}
     { \tilde{\mu}_{\tilde{\sigma}}( \tilde{I}^{(n)}) }\nonumber\\
&=&\lim_{k \rightarrow \infty} 
\frac{1}{\displaystyle 1 + \lim_{n \rightarrow \infty} 
\frac{\tilde{\mu}_{\tilde{\sigma}}( [\tilde{f}_0^{-k}(1),1] 
\cap \tilde{I}^{(n)}) }
     {\tilde{\mu}_{\tilde{\sigma}}([0,\tilde{f}^{-k}_0(1)] 
\cap \tilde{I}^{(n)} ) } }\;. 
\ea
The measure $\tilde{\mu}_{\tilde{\sigma}}( [0,\tilde{f}^{-k}_0(1)] 
\cap \tilde{I}^{(n)})$ can be treated similarly as $\mu(f^{-n}_0(1))$ 
in Equation (\ref{MeasureInx=0}):
\ba
\label{denominator}
\tilde{\mu}_{\tilde{\sigma}}( [0,\tilde{f}^{-k}_0(1)] \cap 
\tilde{I}^{(n)}) \geq 
\tilde{\mu}_{\tilde{\sigma}}(\tilde{I}^{(n)}_0)\hspace{1cm}\nonumber\\
\hspace{1cm}= m\left(\tilde{C}_n(1) e^{-\tilde{\lambda}_0 n} \right)
\tilde{C}_n(1)^{\tilde{\sigma}} e^{-\tilde{\lambda}_0 \tilde{\sigma} n}\;.
\ea
The expression $\tilde{\mu}_{\tilde{\sigma}}( [\tilde{f}_0^{-k}(1),1] 
\cap \tilde{I}^{(n)})$ is the measure of an interval set located 
in $[\tilde{f}^{-k}_0(1),1]$ with total length not greater than 
$e^{-\tilde{\kappa}_1 n}$, 
which is the length of the $n$-th preimage interval set. This measure is
not 
greater than the maximum of the measure of such interval sets. Since for
any fixed value $k$ there exist a 
$\tilde{\mu}_{\tilde{\sigma},\rm max}'(k)$ finite upper bound of
the derivative of the conditionally invariant measure in $[f^{-k}_0(1),1]$,
\begin{equation}
\label{numerator}
\tilde{\mu}_{\tilde{\sigma}} 
\left( [f^{-k}_0(1),1] \cap \tilde{I}^{(n)} \right) 
\leq \tilde{\mu}_{\tilde{\sigma},\rm max}'(k)  e^{-\tilde{\kappa}_1 n}\;.
\end{equation}
Using inequalities (\ref{denominator}), (\ref{numerator}) 
and that $\tilde{\lambda}_0 \tilde{\sigma}<\tilde{\kappa}_1$
due to the criticality, the limit of the fraction in the denominator
of the right hand side of Eq.\ (\ref{ConjugNuin0}) is equal to zero, so 
\begin{equation}
\tilde{\nu}_{\tilde{\sigma}}(\{0\}) = 1 \label{nu0l}
\end{equation} 
whenever $\tilde{\mu}_{\tilde{\sigma}}$ is a critical conditionally 
invariant measure. Now we prove that not only the conjugated natural 
measure, but the original one is concentrated in the fixed point, too. 
We have already seen that for any fixed $k$
\begin{equation}
\label{nu0conv}
\lim_{n \rightarrow \infty} 
\frac{ \tilde{\mu}_{\tilde{\sigma}}( \tilde{I}_0^{(k)} \cap
\tilde{I}^{(n)})}
     { \tilde{\mu}_{\tilde{\sigma}}( \tilde{I}^{(n)}) } = 1\;.
\end{equation}
Since the conjugation does not change the measure of any single interval,
i.e. $\mu_\sigma(I^{(n)}_i) =
\tilde{\mu}_{\tilde{\sigma}}(\tilde{I}^{(n)}_i)$, the
same equation applies for the non-conjugated map, which means 
\begin{equation}
\label{nu0}
\nu_\sigma(\{0\})=1
\end{equation}
for any critical measures.

By using this critical measure as conjugating function one can get the
equivalent Lebesgue map $\tilde f^{(\sigma)}$
where the Lebesgue measure represents the critical state.
The density of its corresponding natural measure is $\delta(x+0)$.
In this Lebesgue map the density of the measure of the coarse grained repeller
$I^{(n)}$ is given by
the $n$-th iterate of $P^{(0)}(x)=1$ by the adjoint
of the Frobenius-Perron equation:
\be
L^+ g=\left\{
\begin{array}{lll}
g(\tilde f^{(\sigma)}(x))&
\mbox{ if }   & \tilde f^{(\sigma)}(x) \in [0,1]\;,\\
0& \mbox{ if }& \tilde f^{(\sigma)}(x) \not \in [0,1]\;.
\end{array}
\right.
\ee
This equation has $\delta(x+0)$ as eigenfunction with eigenvalue
$e^{\kappa\sigma}=\left.\frac{d}{dx}\tilde f^{(\sigma)}(x)\right|_{x=0}$.
Our result (\ref{nu0l}) amounts to proving that
${L^+}^n P^{(0)}(x)$
converges to $\delta(x+0)$ when $n\rightarrow\infty$.
This convergence property has been assumed previously supported by numerical
calculations and also some of its
consequences have been exploited \cite{NSz,LSz,KLNSz}.

From Equation (\ref{nu0}) follows that $\lambda_0=\lambda$,
where $\lambda$ is the average Liapunov exponent,
and the Kolmogorov-Sinai entropy $K$ is zero
for the natural measure in the critical case.
The equations
\be
\kappa=\sigma\lambda\;,\;\; K=0
\ee
valid for the critical states
are the counterparts of the generalized Pesin relation
\be
\kappa=\lambda-K \label{kg}
\ee
valid for noncritical states, obtained by Kantz and Grassberger
\cite{KG,ER}.

Finally we note, that
the map $f$ can be considered to be the reduced map of a translationally
invariant map of the real axis \cite{GeNi,SFrKa,GrFu,Ga,KlD99}.
Then the diffusion coefficient can be written as an average over the natural
measure of the reduced map \cite{KLNSz}.
This in case of critical state obviously results in a zero diffusion
coefficient.
In the noncritical state there are important connections
between the diffusion and the formula (\ref{kg}) \cite{GaN,TVB,K99}.

\section{Examples}

In this section we demonstrate the properties we have found along with further
discussion.
As an example consider the map whose inverse branches are
\begin{eqnarray}
\label{LebInv}
f^{-1}_0(x) & = & \frac{1+d}{2R}x-\frac{d}{4R^2}x^2\;, \nonumber\\
f^{-1}_1(x) & = & 1- \frac{1-d}{2R}x-\frac{d}{4R^2}x^2\;,
\end{eqnarray}
where $R>1$ and $-1<d\leq 1$ must hold.
The case $d=0$ corresponds to the case of the tent map and the eigenvalue
$\Lambda_\sigma=(2R)^{-\sigma}$ has been already given by \cite{MN}.
Eq.\ (\ref{kl}) for $d=0$ shows in which region one can connect this
eigenvalue with the escape rate.
The map is conjugated to the symmetric piecewise parabolic map
\cite{GySz/Z,NSz,LSz,KLNSz}.
For the sake of simplicity
we limit our discussion to non-negative values of $d$.
Substituting the inverse branches into the Frobenius-Perron equation, 
one can immediately see that the Lebesgue measure is a conditionally 
invariant measure with the escape rate $\kappa_1=\log R$, independently 
of the value of $d$. Similarly, the exponent $\psi$ is
equal to unity for any $0 \leq d \leq 1$.
However, $\omega$ and consequently $\beta=\psi\omega$ have
two possible values depending on $d$.
This makes it sensible to analyse this 
map in two parts, according to the value of $\beta$. 
Let us start with $0 \leq d <1$, when $\beta=\omega=1$.
The value 
of $\kappa_\beta=\kappa_1=\log R$ is exactly known, therefore
\begin{equation}
\sigma_c = \frac{\kappa_\beta}{\lambda_0} = 
\frac{\log R}{\log R+\log\frac{2}{1+d}}\;.
\end{equation}
Numerical results for $\kappa_\sigma$ fit to Eqs.\ (\ref{ks1}), (\ref{ks2}),
as seen in Fig.~2.
Fig.\ 3 shows some of the numerically obtained
conditionally invariant densities.

In the case $d=1$ obviously $\beta=\omega=2$.
Then $\kappa_\beta$ is not known 
exactly, but it can be determined numerically. Numerical
calculation for $R=1.5$ gave $\kappa_\beta\approx 0.60$ and 
$\sigma_c = \kappa_\beta/\log R \approx 1.48$.
Accordingly to the results of Section 2
conditionally invariant measures were found for $\sigma<\sigma_c$
and values of $\kappa_\sigma$ fit to Eqs.\ (\ref{ks3}), (\ref{ks4})
(see Figs.~2 and 3).
However, critical slowing down of convergence is seen near $\sigma_c$.

\begin{figure}
\epsfbox{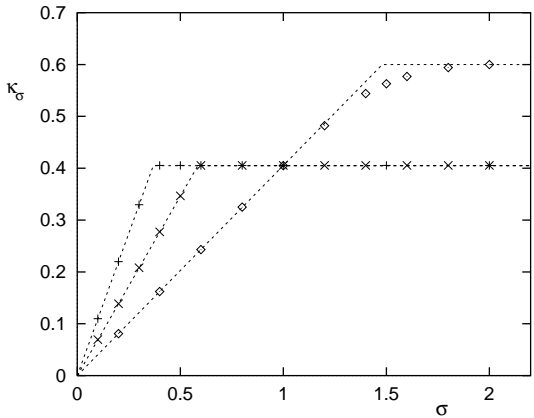}
\vspace*{0.5mm}
\caption{Numerical results for $\kappa_\sigma$ in case of the map
(\ref{LebInv}) with $R=1.5$ and
$d=0$ (+), $d=0.5$ ($\times$), $d=1$ ($\Diamond$);
and the theoretical results (\ref{ks1}), (\ref{ks2}), (\ref{ks3}),
(\ref{ks4}) (dashed lines) for the same values of $d$.}
\end{figure}

\begin{figure}
\epsfbox{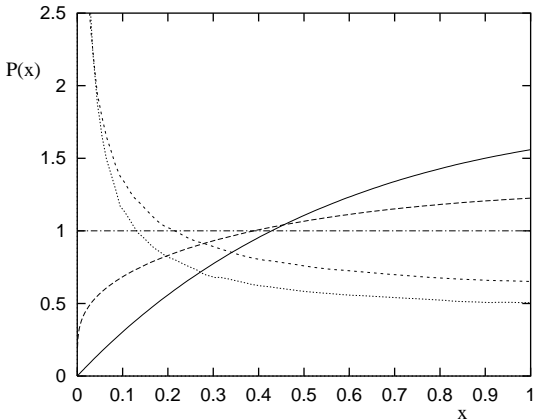}
\vspace*{0.5mm}
\caption{The densities of the conditionally invariant measures of the map
(\ref{LebInv}) with $d=0, \sigma=0.2$ (dotted line);
$d=0.5, \sigma=0.4$ (short dashes);
$d=1, \sigma=1.2$ (long dashes);
$d=1, \sigma=2$ (solid line);
and the constant density of the Lebesgue measure obtained numerically
for $\sigma=1$ and all the three values of $d$ (dashed dotted line).}
\end{figure}

Another map was constructed for that $\psi=1$ and $\beta=\omega=4$.
Its inverse branches are
\begin{eqnarray}
f^{-1}_0(x) &=& \frac{x}{R}-\frac{x^4}{Q}\;,\nonumber\\
f^{-1}_1(x) &=& 1-\frac{x^4}{Q}\;,\label{4inv}
\end{eqnarray}
where $R>1$ and $Q\ge 4R$.
In the numerical calculations $R=1.25$ and $Q=40$ was used.
The Lebesgue measure is again one of the conditionally invariant measures
with escape rate $\kappa_1=\log R$.
From the numerical value $\kappa_\beta\approx 0.730$ follows that
$\sigma_c=\kappa_\beta/\log R \approx 3.27$.
Numerical values of $\kappa_\sigma$ are compared to the theoretical values
in Fig.~4.
Presence of the conditionally invariant measure that is smooth at least
in the inside of $[0,1]$ was checked numerically at several values of
$\sigma$ with $\sigma<\sigma_c$ and at $\sigma=\beta$.
Among them the ones with integer $\sigma$ have analytic leading term 
at $x=0$.
The densities of the latter ones together with a singular one are seen
in Fig.~5.

\begin{figure}
\epsfbox{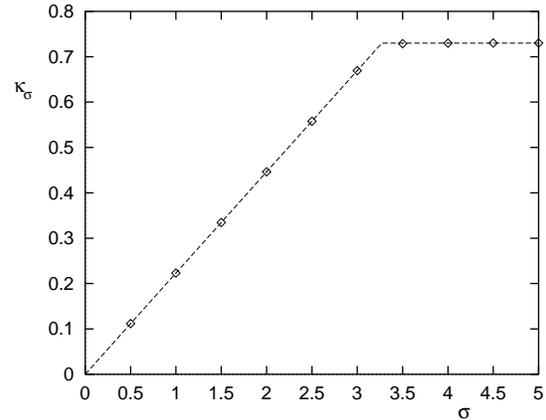}
\vspace*{0.5mm}
\caption{Numerical ($\Diamond$) and theoretical (dashed line) results
for $\kappa_\sigma$ in case of the map (\ref{4inv}).}
\end{figure}

\begin{figure}
\epsfbox{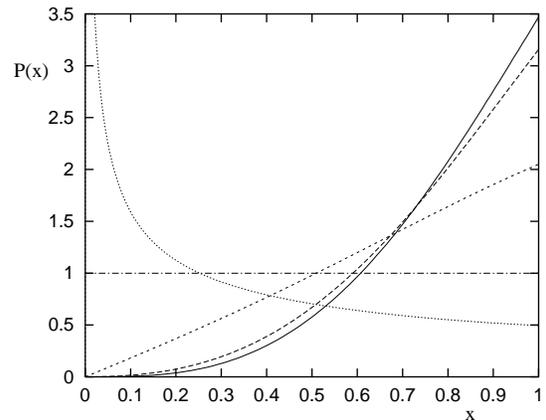}
\vspace*{0.5mm}
\caption{The conditionally invariant densities of the map (\ref{4inv})
with $\sigma=0.5$ (dotted line), $\sigma=1$ (dashed dotted line),
$\sigma=2$ (short dashes), $\sigma=3$ (long dashes),
and $\sigma=4$ (solid line).}
\end{figure}

\acknowledgments

This work has been supported in part
by the Hungarian National Scientific Research Foundation under Grant 
Nos.\ OTKA T017493 and OTKA F17166.

\end{document}